\documentclass[a4paper,11pt]{article}

\usepackage{amsmath, amssymb, bm}
\usepackage[]{graphicx}
\usepackage[caption=false]{subfig}
\usepackage{fullpage}
\usepackage{authblk}
\usepackage{url}
\usepackage{hyperref}

\begin{document}

\title{\bf{Axion production and CMB spectral distortion in cosmological tangled magnetic field}}
\author[]{Damian Ejlli}
\affil{\emph{Theory group, INFN Laboratori Nazionali del Gran Sasso, 67100 Assergi, L'Aquila Italy and Department of Physics, Novosibirsk State University, Novosibirsk 630090, Russia }}

\date{}

\maketitle

\begin{abstract}

Axion production due to photon-axion mixing in tangled magnetic field(s) prior to recombination epoch and magnetic field damping can generate cosmic microwave background (CMB) spectral distortions. In particular, contribution of both processes to CMB $\mu$ distortion in the case of resonant photon-axion mixing is studied. Assuming that magnetic field power spectrum is approximated by a power law $P_B(k)\propto k^n$ with spectral index $n$, it is shown that for magnetic field cut-off scales $172.5$ pc $\leq \lambda_B\leq 4\times 10^3$ pc, axion contribution to CMB $\mu$ distortion is subdominant in comparison with magnetic field damping in the cosmological plasma. Using COBE upper limit on $\mu$ and for magnetic field scale $\lambda_B\simeq 415$ pc, weaker limit in comparison with other studies on the magnetic field strength ($B_0\leq 8.5\times 10^{-8}$ G) up to a factor 10 for the DFSZ axion model and axion mass $m_a\geq 2.6\times 10^{-6}$ eV is found. A forecast  for the expected sensitivity of PIXIE/PRISM on $\mu$ is also presented.

\end{abstract}

\url{damian.ejlli@lngs.infn.it}

\vspace{1cm}

\emph{Introduction}. During the last decades have been done intensive studies regarding the existence and nature of primordial magnetic field(s) at both small and large scales. Its existence could have strong impact in different cosmological scenarios such as bing bang nucleosynthesis (BBN), structure formation, CMB temperature anisotropy etc. In general, in all those scenarios, it is possible to probe its existence only indirectly, namely through the coupling of magnetic field with the cosmological plasma. Consequently, based on information that we have on BBN, CMB temperature anisotropy etc., it is possible to speculate about the magnetic field structure and estimate its strength at a given scale. In particular, CMB temperature anisotropy has been one of the most important benchmark to test the existence of  primordial magnetic field(s). Indeed, an ubiquitous, anisotropic and homogeneous magnetic field with strength at present time $B_0\lesssim 3\times 10^{-9}$ G would create the observed CMB temperature anisotropy due to anisotropic expansion of the Universe \cite{zel'dovich70}. For a general review on cosmological magnetic field see Refs. \cite{Grasso:2000wj}.

In the presence of a large scale magnetic field, CMB photons can in principle convert into axions or other similar particles due to their coupling with the magnetic field. In Refs. \cite{Ejlli:2013uda} we have studied such a mechanism in the presence of large scale \emph{uniform} (spatially homogeneous) magnetic field and applied it to CMB spectral distortions and temperature anisotropy. However, several interesting questions arises as to what happens in the case when the background magnetic field is not homogeneous (tangled magnetic field). Does the magnetic field has an impact on the spectral distortions? Is the impact of the magnetic field on the spectral distortions dominant or subdominant with respect to photon-axion oscillations? 

As in the case of density perturbations in the primordial baryonic plasma suffering from Silk damping, we can expect that a spatially varying magnetic field can couple to the baryon plasma and dissipate energy. This would eventually lead to the damping of the primordial magnetic field spectrum in different scales \cite{Jedamzik:1996wp}.  In general, a distinguishing feature of non homogeneous magnetic field in comparison with an uniform field is that the former can have an impact on the CMB by distorting its spectrum. Indeed, it has been shown in Ref. \cite{Jedamzik:1999bm} that a spatially varying stochastic magnetic field may significantly dissipate in the cosmological plasma prior to recombination epoch. By dissipation, the magnetic field energy would transform into kinetic energy of cosmological plasma and in turn plasma's kinetic energy would be efficiently transformed into heat due to high shear viscosity of plasma. In the limit when photon mean free path $l_\gamma$ is smaller than magnetic field mode $\lambda$, $l_\gamma \ll \lambda$, Alfv\'{e}n, slow and fast magnetosonic  waves with $\lambda<d_\gamma$ are effectively dissipated where $d_\gamma=(l_\gamma\,t)^{1/2}$ is the photon diffusion length and $t$ is the cosmological time.

In the case when there is an energy injection into the cosmological plasma such as conversion of magnetic field energy into heat, electrons gain energy and the electron temperature $T_e$ becomes higher than the photon temperature $T$, $T_e>T$. Depending at which redshift the magnetic field energy is converted into heat, this effect would eventually lead to CMB spectral distortions if energy is injected for redshift $z\lesssim 2\times 10^6.$ For an early treatment of CMB spectral distortion see Ref. \cite{Zeldovich:1969ff}, for further developments see Ref. \cite{CMB-spectral} and for production mechanisms of spectral distortions see Ref. \cite{Hu:1993gc}. 

\begin{figure*}[htbp!]
\centering
\mbox{
\subfloat[\label{fig:Fig1}]{\includegraphics[scale=0.85]{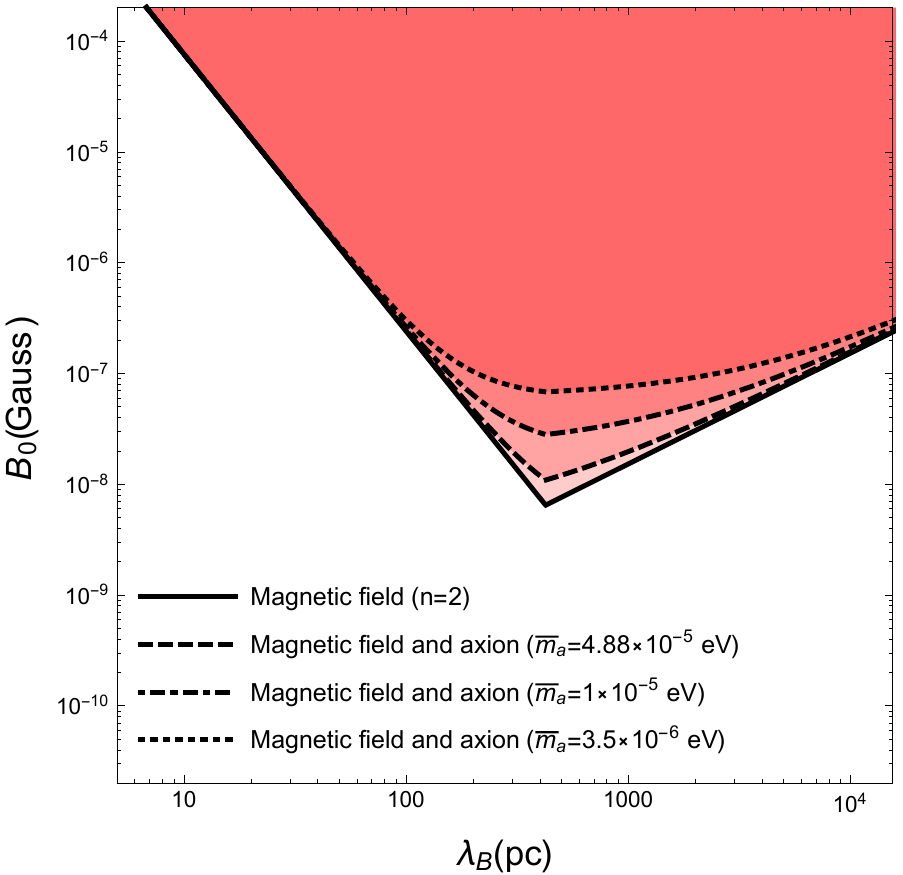}}\qquad
\subfloat[\label{fig:Fig2}]{\includegraphics[scale=0.85]{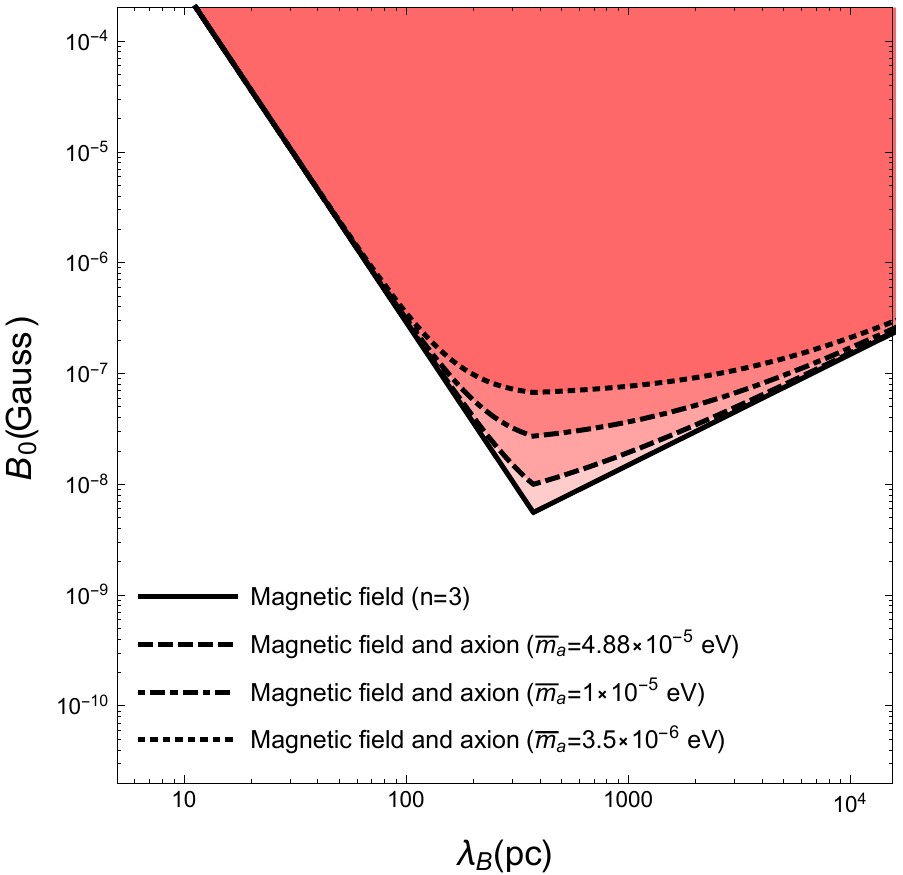}}}
\caption{Exclusion plot in the parameter space  $B_0-\lambda_B$ in the resonant case due to $\mu$-distortion. In (a) the exclusion plot for COBE \cite{Fixsen:1996nj} upper limit on $\mu$ and DFSZ axion model for $n=2$ is shown and in (b) the exclusion plot for COBE upper limit on $\mu$ and DFSZ axion model for $n=3$ is shown. In both (a) and (b) the region above the solid line represents the excluded region without  photon-axion mixing while the region above the dashed, dot dashed and dotted lines represent the exclude region including photon-axion mixing for $\bar m_a=4.88\times 10^{-5}$ eV, $\bar m_a=1\times 10^{-5}$, $\bar m_a=3\times 10^{-6}$ respectively.}
\label{fig:Fig1a}
\end{figure*}

\begin{figure*}[htbp!]
\centering
\mbox{
\subfloat[\label{fig:Fig3}]{\includegraphics[scale=0.85]{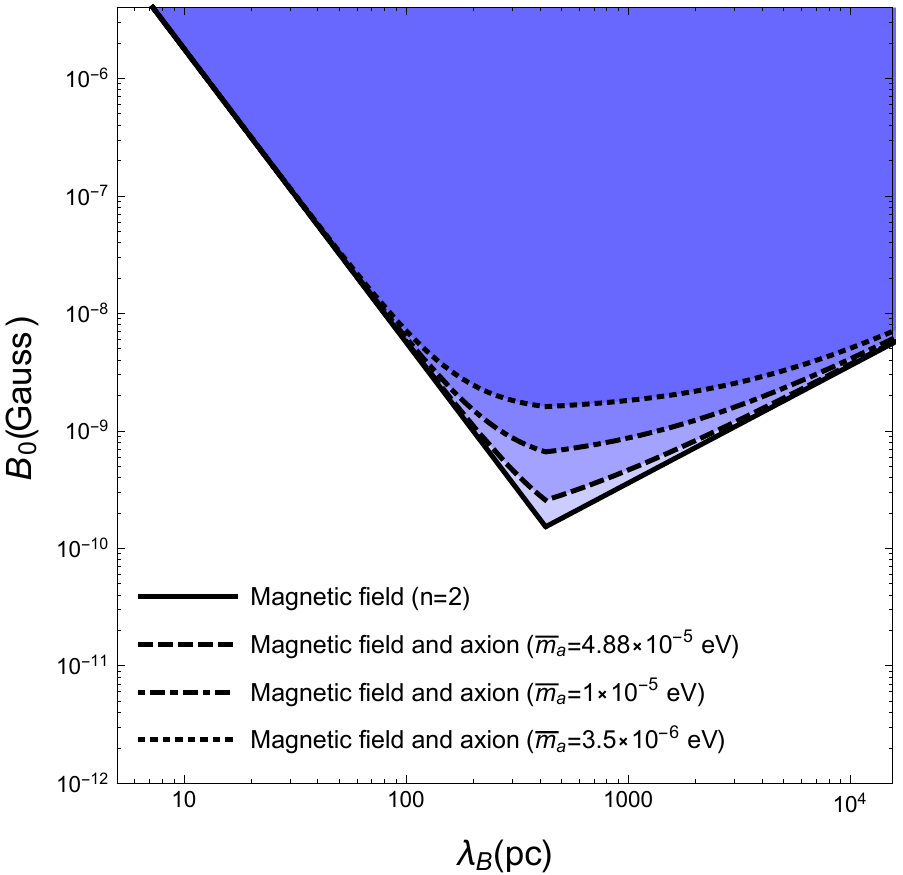}}\qquad
\subfloat[\label{fig:Fig4}]{\includegraphics[scale=0.85]{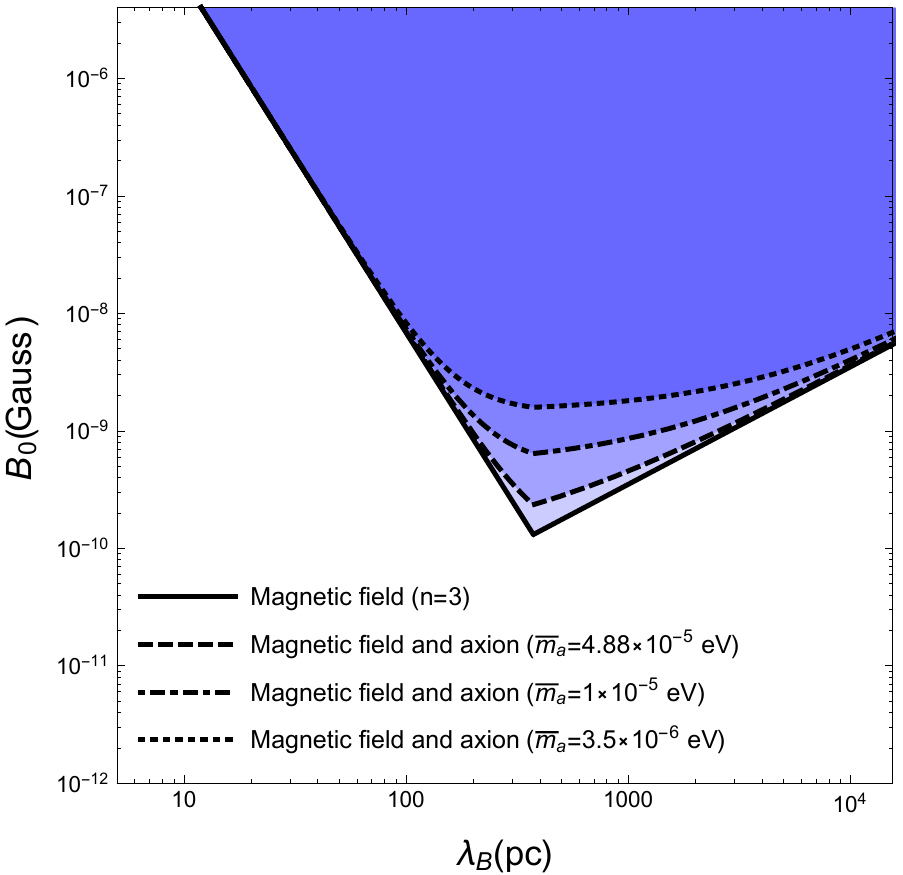}}}
\caption{Sensitivity plot in the parameter space  $B_0-\lambda_B$ for PIXIE/PRISM \cite{Kogut:2011xw} expected value on $\mu\simeq 5\times 10^{-8}$ and for the DFSZ axion model. Values of magnetic field spectral index $n$ and axion mass $\bar m_a$ are the same as in Fig. \ref{fig:Fig1a}.}
\label{fig:Fig2a}
\end{figure*}

\begin{figure*}[htbp!]
\centering
\mbox{
\subfloat[\label{fig:Fig5}]{\includegraphics[scale=0.85]{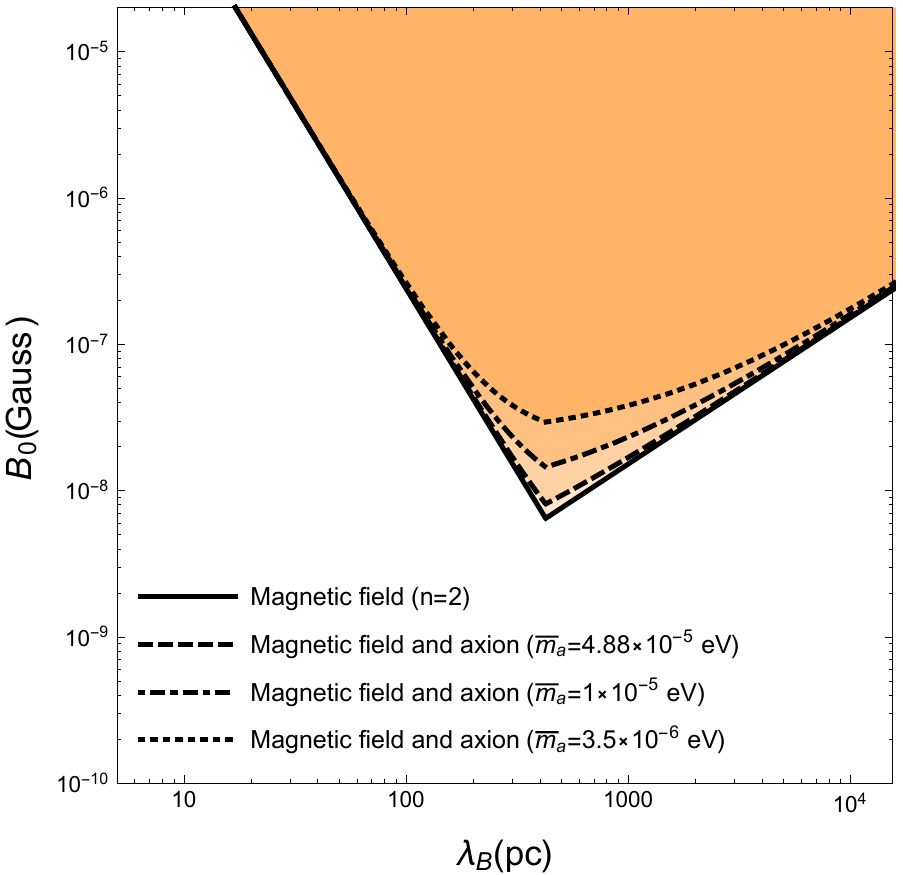}}\qquad
\subfloat[\label{fig:Fig6}]{\includegraphics[scale=0.85]{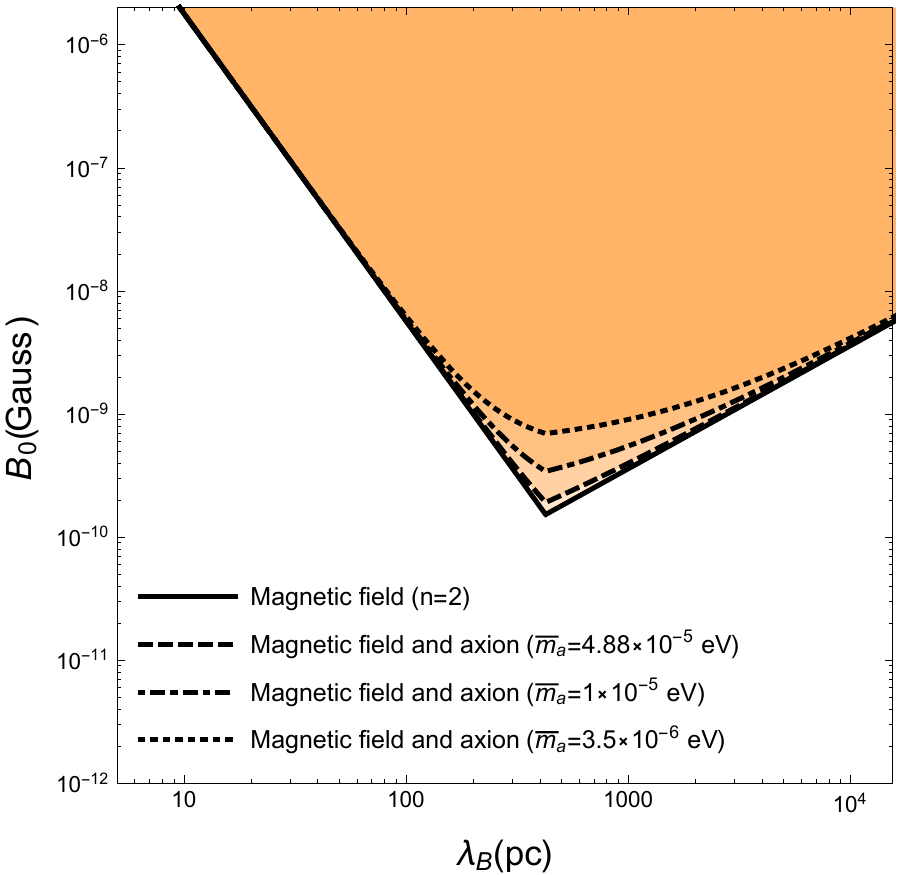}}}
\caption{In (a) the exclusion plot in the parameter space  $B_0-\lambda_B$ for COBE limit on $\mu< 9\times 10^{-5}$ for the KSVZ axion model and $n=2$ is shown and in (b) the sensitivity plot in the parameter space  $B-\lambda_B$ for PIXIE/PRISM expected value on $\mu\simeq 5\times 10^{-8}$ for the KSVZ axion model and $n=2$ is shown. Values of the axion mass are same as in Fig. \ref{fig:Fig1a}}
\label{fig:Fig3a}
\end{figure*}

\begin{figure*}[htbp!]
\centering
\mbox{
\subfloat[\label{fig:Fig7}]{\includegraphics[scale=0.85]{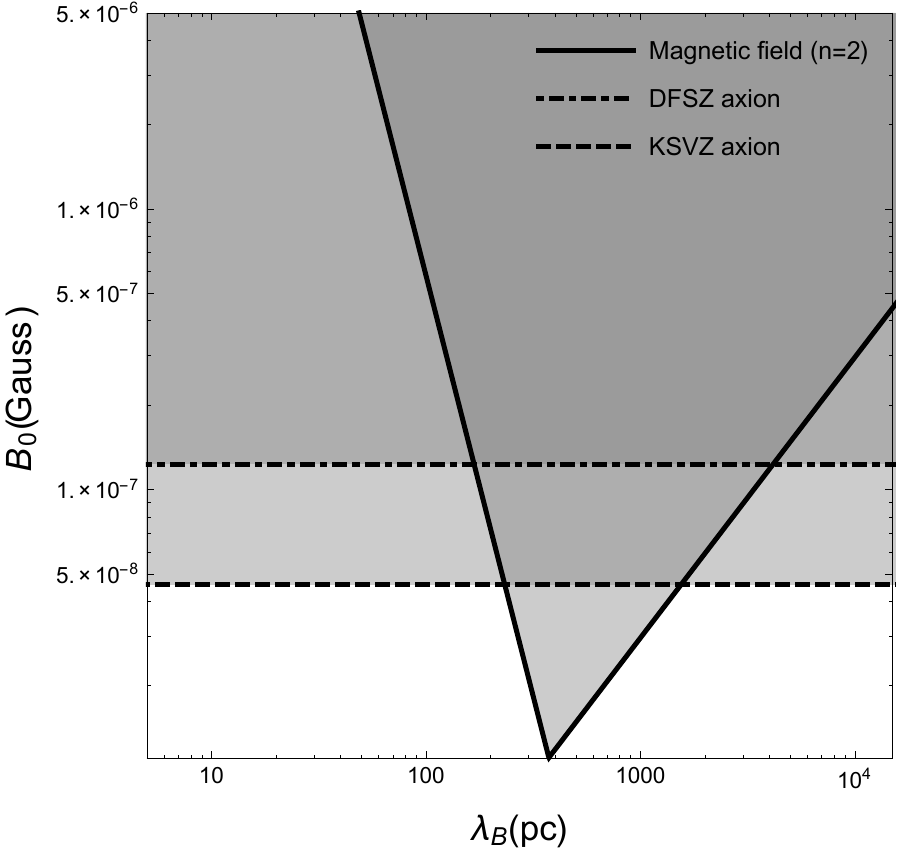}}\qquad
\subfloat[\label{fig:Fig8}]{\includegraphics[scale=0.85]{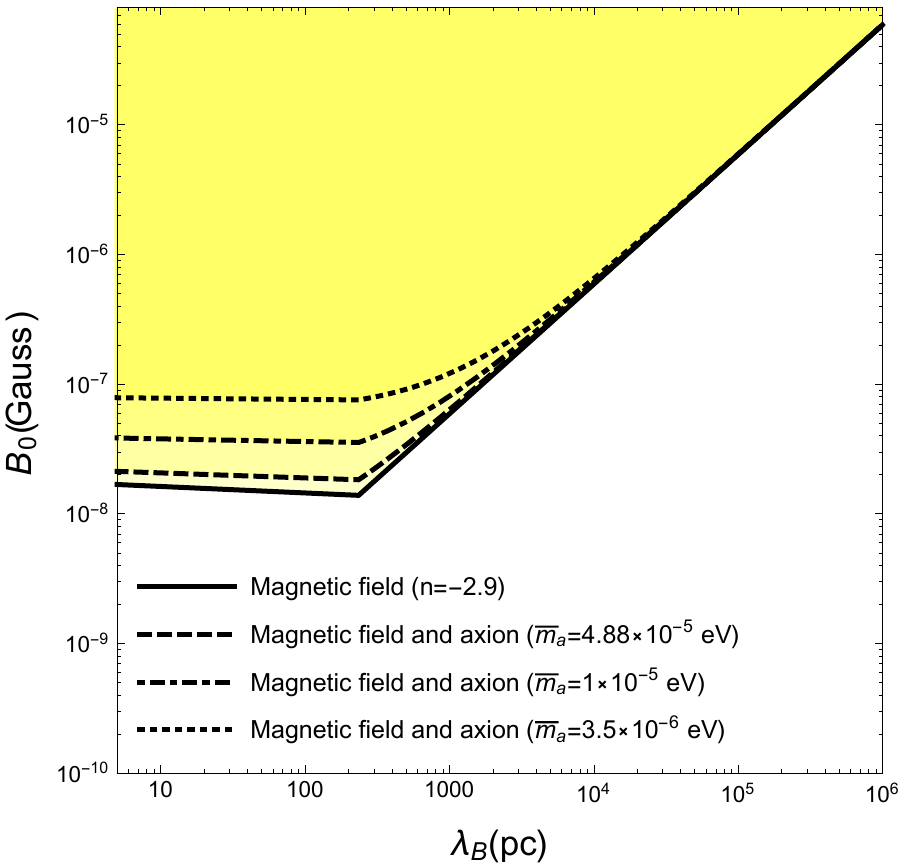}}}
\caption{Exclusion plot in the parameter space  $B_0-\lambda_B$ for COBE limit on $\mu< 9\times 10^{-5}$. In (a) the region above the solid line is excluded with no photon-axion mixing included while the regions above dashed and dot dashed are respectively excluded by KSVZ and DFSZ axion models only. In (b) exclusion plot for magnetic field spectral index $n=-2.9$ and DFSZ axion model is shown. Values of the axion mass are same as in Fig. \ref{fig:Fig1a}.}
\label{fig:Fig4a}
\end{figure*}

\begin{figure*}[htbp!]
\centering
\mbox{
\subfloat[\label{fig:Fig9}]{\includegraphics[scale=0.85]{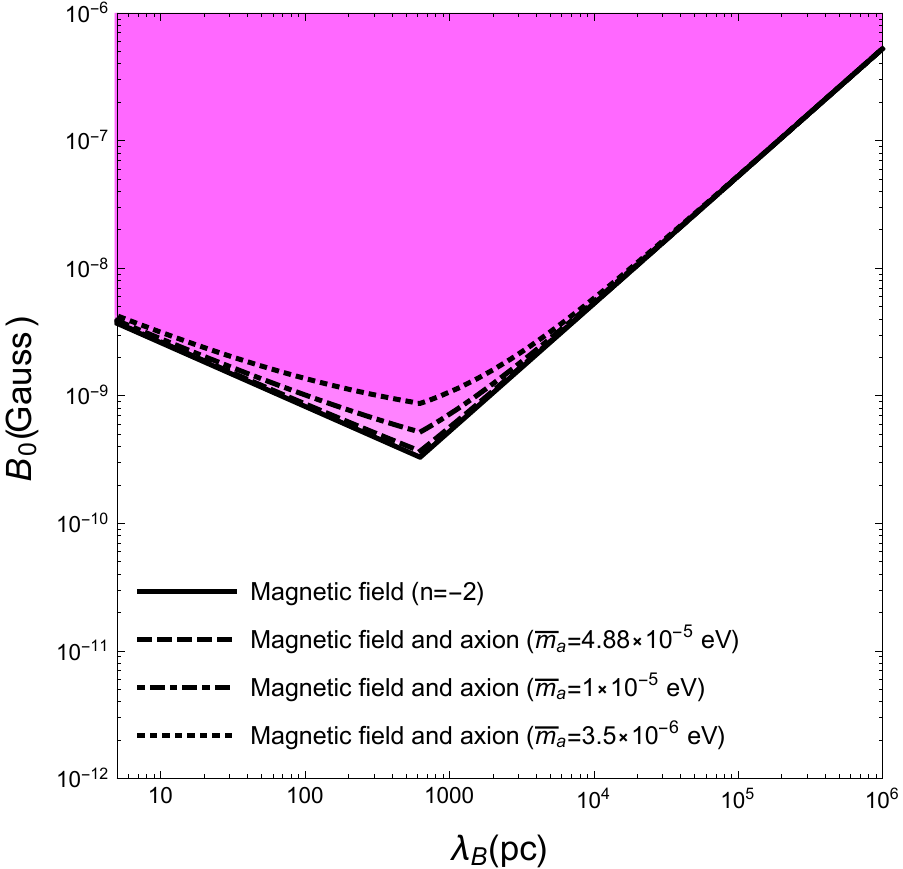}}\qquad
\subfloat[\label{fig:Fig10}]{\includegraphics[scale=0.85]{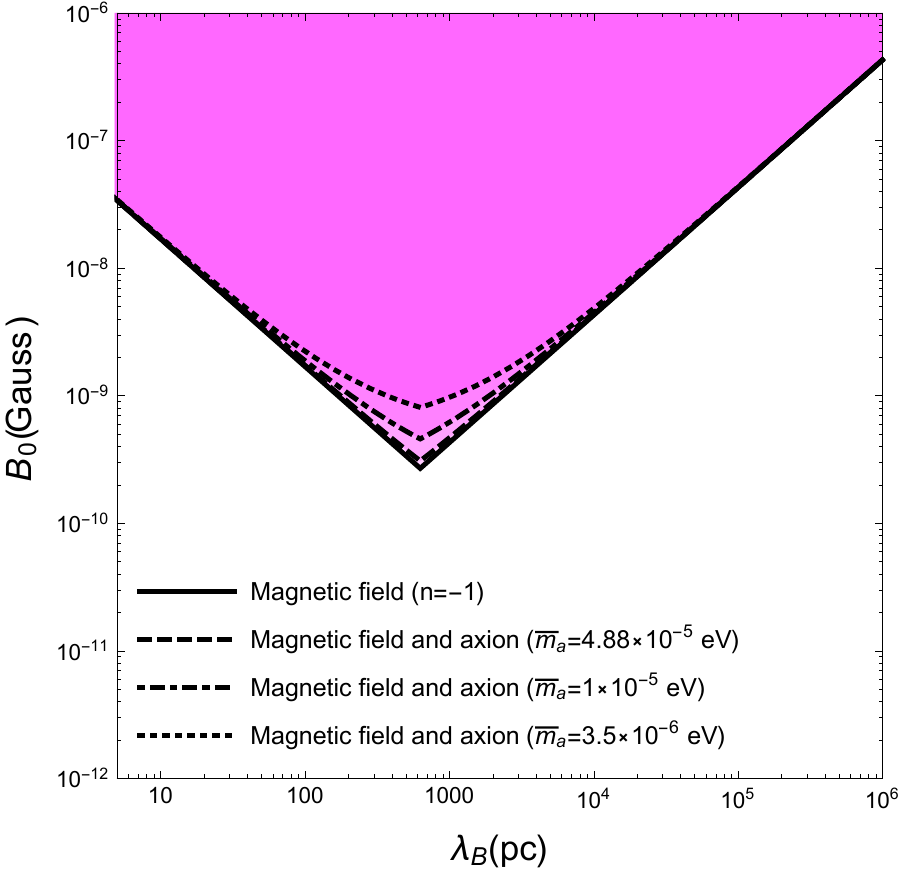}}}
\caption{Sensitivity plot in the parameter space  $B_0-\lambda_B$ for PIXIE/PRISM expected limit on $\mu\simeq 5\times 10^{-8}$. In (a) the sensitivity plot for KSVZ axion model and $n=-2$ is shown while in (b) the sensitivity plot for KSVZ axion model and $n=-1$ is shown.}
\label{fig:Fig5a}
\end{figure*}

\emph{Dissipation of tangled magnetic field}. During the evolution of the universe, it is usually assumed that the conductivity of the cosmological plasma is infinite. In this case the field amplitude scales as $B\sim B_0 a^{-2}(t)$ where $a$ is the cosmological scale factor. Even though this is a good approximation, it does not reflect the more general case, namely in the case of tangled magnetic fields when the magnetic field dissipate energy. In order to make contact with our results that follow, we assume that the magnetic field is generated by random processes (stochastic) in the early universe during inflation or radiation epoch and it evolves according to the following law
\begin{equation}
\mathbf B(\mathbf x, t)=\mathbf b_0(\mathbf x, t)\left(\frac{a_0}{a}\right)^2,
\end{equation}
where $a_0$ is the cosmological scale factor at present epoch and $\mathbf b_0(\mathbf x, t)$ is the tangled magnetic field of which amplitude evolves as $b=b_0\exp(i\int_{t_0}^t dt^\prime \omega)$ where $\omega$ is the magnetic field mode frequency, see Refs. \cite{Jedamzik:1996wp} and \cite{Jedamzik:1999bm}. Moreover we assume that magnetic field is \emph{statistically} homogeneous and isotropic with ensemble average
\begin{equation}
\langle b_i(\mathbf k)b_j^*(\mathbf q)\rangle=\delta^3(\mathbf k-\mathbf q)P_{ij}(\mathbf k)P_B(k),
\end{equation}
where $P_{ij}$ is a projection tensor and $P_B$ is the power spectrum of primordial magnetic field that in general is assumed to be a power law, $P_B=C k^{n}$ with $C$ a constant and $n$ the spectral index of the magnetic field. The constant $C$ is fixed by taking the spatial average (or ensemble average) of the energy density of the magnetic field over a volume $V$
\begin{equation}
\rho_B(t_0)=\frac{\langle \mathbf B_0(\mathbf x)\mathbf B_0(\mathbf x)\rangle}{2}=\frac{B_0^2}{2},
\end{equation}
with comoving cut-off wavelength $\lambda_B=2\pi/k_\lambda$ where $k_\lambda$ is a cut-off comoving wave-vector. The cut-off wavelength (or wave-vector) is in general a free parameter that is connected with Fourier decomposition of $\bf B(\bf x)$ and in principle can assume values from zero to infinity. However, for physical reasons it depends essentially on the generation mechanism of the primordial magnetic field. In the case of magnetic field generated by casual mechanism(s), the value of $\lambda_B$ should be smaller or equal to the Hubble horizon while in the case of magnetic field generated by non casual mechanism(s) (in general negative spectral indexes) the value of $\lambda_B$ can be greater than Hubble horizon. As we see below, in the case of casual mechanisms, we consider that $\lambda_B$ to be smaller or equal to Hubble horizon during the $\mu$ epoch, see Figs. \ref{fig:Fig1a}, \ref{fig:Fig2a} and \ref{fig:Fig3a}. In the case of magnetic fields generated by non casual mechanisms we set for simplicity $\lambda_B$ to be of the order of Mpc even though it can be larger than this value, see Figs. \ref{fig:Fig8} and \ref{fig:Fig5a}. 

With this kind of normalisation the magnetic field power spectrum is given by
\begin{equation}\label{B-power-spec}
P_B=\frac{B_0^2}{4\pi}(n+3)\left(\frac{k}{k_\lambda}\right)^n.
\end{equation}
Another possibility on fixing the constant $C$, that is also used in the literature, is to use a Gaussian filter $e^{-(k/k_\lambda)^2}$ in the definition of $\rho_B(t_0)$, see Refs. \cite{Kunze:2013uja}. In this case the form of $P_B$ is different from Eq. \eqref{B-power-spec} but the spatial average of $\rho_B(t_0)$ remains invariant as it should be. In this work we shall not adopt this definition.

Let $\dot Q_B$ indicate the energy loss per unit time of magnetic field that would convert into heat in the plasma. If energy injection (or heat) occurs in the redshift interval $2.88\times 10^5\leq z\leq 2\times 10^6$, the Compton scattering would eventually create a Bose-Einstein distribution for the photon spectrum with chemical potential $\mu$\footnote{The chemical potential introduced here is an dimensionless quantity and is related to the thermodynamical chemical potential $\mu_{ther}$ by $\mu=-\mu_{ther}/T.$}. The evolution of chemical potential with respect to time is governed by the Sunyaev-Zel'dovich equation \cite{Zeldovich:1969ff}
\begin{equation}\label{chem-pot}
\frac{d\mu}{dt}=-\frac{\mu}{t_{dC}}+1.4\,\frac{\dot Q_B}{\rho_R},
\end{equation}
where $t_{dC}=2.09\times 10^{33}(1-Y_p/2)^{-1}(h^2\Omega_B)^{-1}(1+z)^{-9/2}$ s is the characteristic time for double Compton scattering. Here $Y_p\simeq 0.24$ is the helium primordial weight by mass and $h^2\Omega_B\simeq 0.022$ is the density parameter of baryons \cite{Ade:2013zuv}.

To solve Eq. \eqref{chem-pot} we need to know the rate of heat flow into the plasma due to magnetic field dissipation. It can be shown that in the photon diffusion limit i.e. $\Gamma_\gamma^{-1}\ll \lambda_B$ \cite{Jedamzik:1996wp} (where $\Gamma_\gamma=\sigma_T n_e$ with $\sigma_T$ being the Thomson cross section and $n_e$ the number density of free electrons)
\begin{equation}\label{eq-heat}
\frac{\dot Q_B}{\rho_R}=\frac{B_0^2}{2\rho_R(t_0)}\frac{(n+3)}{k_\lambda^{n+3}}\int_0^{k_\lambda}\,dk\frac{k^{n+4}}{5(1+z)\Gamma_\gamma(t_0)}\exp{\left(-\frac{2k^2}{k_D^2(t_0)}\frac{1}{(1+z)^3}\right)},
\end{equation}
where $\rho_R$ is the energy density of relativistic particles, $t_0$ denotes the present time and $k_D^2(t_0)=15\,\Gamma_\gamma(t_0)/t_*\simeq 6.27\times 10^{-19}\Gamma_\gamma(t_0)$ s$^{-1}$. The term $t_*$ is connected to the thermalization redshift\footnote{The thermalization redshift is the redshift that for $z\geq z_{\mu}$ the CMB spectrum is in thermal equilibrium and for $z<z_{\mu}$ the spectrum is a Bose-Einstein distribution, see Refs. \cite{CMB-spectral}.}, $z_\mu$, through the relation $t_*=5t_{dC}/(4(1+z_\mu)^{5/2})$. Substituting Eq. \eqref{eq-heat} into Eq. \eqref{chem-pot}, the general solution of Eq.  \eqref{chem-pot} in terms of the redshift is given by
\begin{equation}\label{red-shift}
\mu(z)=\frac{1.4\, B_0^2}{10\, \rho_R(t_0)}\frac{(n+3)}{\Gamma_\gamma(t_0)\,k_\lambda^{n+3}}\int_{0}^{k_\lambda}\int_z^{z_i}dz^\prime dk\,\frac{k^{n+4}}{(1+z^\prime)^4}\exp{\left(-\frac{1+z^\prime}{1+z_\mu}\right)^{5/2}}\exp{\left(-\frac{2k^2}{k_D^2(t_0)(1+z^{\prime })^3}\right)},
\end{equation}
where $z_i$ is an initial redshift, $z_i\gg z_\mu$ and the term proportional to $\mu(z_i)$ is absent since for $z_i$ we have $\mu(z_i)=0.$ In obtaining Eq. \eqref{red-shift} we have used the fact that in the radiation dominated universe $dt=-dz\,2 t_*/(1+z)^3$ and $z$ is the redshift in the radiation dominated universe $z\ll z_\mu.$

In general is not possible to find analytic solution of Eq. \eqref{red-shift} due to the non trivial form of the integrands. Indeed, one can recognise that the double integral of Eq. \eqref{red-shift} can be expressed in terms of the incomplete gamma functions, $\gamma(s, 2k_\lambda^2/k_D^2)$ where $s$ is an integer that in our case is either $s=(n+5)/2$ or $s=(3n+9)/5$. However, it is possible to consider some limiting cases that allows to find analytic expressions in terms of Euler gamma function $\Gamma$. Let us consider the limit $k_\lambda^2\gg k_D^2(t_0)(1+z_\mu)^3$ and then evaluate the residual chemical potential at redshift $z=0$ (today). In this limit we get the following relation between magnetic field strength $B_\textrm{0}$ and $\mu$
\begin{equation}\label{heat-B-limit}
B_0=3.19\times 10^{-6}\sqrt{\frac{\mu}{C_n}}\left(\frac{k_D}{k_\lambda}\right)^{-(n+3)/2}\,\textrm{G},
\end{equation}
where $C_n$ is a constant 
\begin{equation}
C_n=1.4\,\Gamma(n/2+5/2)\,\Gamma(3n/5+9/5)\,2^{-(n+5)/2}\,(6/5)\,(n+3),
\end{equation}
Here the term $k_D=k_D(t_0)z_\mu^{3/2}$ in Eq. \eqref{heat-B-limit} is the scale damped by one e-fold at redshift $z_\mu.$ Its corresponding co-moving wavelength is $\lambda_D=2\pi/k_D=415.5$ pc. On the opposite, in the limit  $k_\lambda^2\ll k_D^2(t_0)(1+z_\mu)^3$, we get
\begin{equation}\label{limit-2}
B_0=3.19\times 10^{-6} \sqrt{\frac{\mu}{D_n}}\left(\frac{k_D}{k_\lambda}\right)\,\textrm{G},
\end{equation}
where $D_n$ is a numerical constant that is given by
\begin{equation}
D_n=1.4\,\Gamma(-6/5)\left(\frac{n+3}{n+5}\right)(6/5).
\end{equation}

\emph{Axion contribution to spectral distortion}. We have seen that tangled magnetic fields can dissipate energy and create $\mu$ distortion in the early universe. However, their presence would make possible the transition of CMB photons into axions\footnote{In this paper we focus only on the QCD axion (hadronic axions)}. In Refs. \cite{Ejlli:2013uda} we have derived the equations of motions for the photon-axion system in the steady state approximation in the case of uniform magnetic field. Here we calculate the transition probability in the resonance case in presence of tangled magnetic field. The resonant regime is the axion mass range that makes resonant transition in the redshift interval $2.88\times 10^5\lesssim 1+z\lesssim 2\times 10^6$, see Ref. \cite{Ejlli:2013uda} for details. The difference in this case is that the magnetic field depends on the position, $\mathbf B(\mathbf x, t)$. However, one does not need to calculate the equation of motion for the density operator $\hat\rho$ again. It is only sufficient to replace in the equations of motions for $\hat\rho$,  $\mathbf B(t)\rightarrow \mathbf B(\mathbf x, t)$ and take the spatial average of the transition probability, $P_a\rightarrow \langle P_a\rangle$. Since the transition probability depends on $\mathbf B_0^2$ \cite{Ejlli:2013uda} and using the fact that  $\langle \mathbf B_0(\mathbf x)\mathbf B_0(\mathbf x)\rangle=B_0^2$, in the resonant case we get 
\begin{equation}\label{res-probab}
\langle P_a(\bar T)\rangle=5.75\times 10^{-27}\,x\,C_{a\gamma}^2\,B_\textrm{nG}^2 \left(\frac{\bar T}{T_0}\right)^3,
\end{equation}
where $B_\textrm{nG}=(B_0/\textrm{nG})$ and $C_{a\gamma}$ is defined as 
\begin{equation}
C_{a\gamma}\equiv  \left(\frac{E}{N}-\frac{2}{3}\frac{4+w}{1+w}\right)\frac{1+w}{w^{1/2}},
\end{equation}
where for $w=0.56$, $|C_{a\gamma}|\simeq 4$ for $E/N=0$ (KSVZ axion model) and $|C_{a\gamma}|\simeq 1.49$ for $E/N=8/3$ (DFSZ axion model). Here $w$ is defined in terms of the mass ratio of up and down quarks, $w=m_u/m_d$. For small chemical potential $\mu$ we can write $\langle P_a\rangle=\mu e^x/(e^x-1)$ where $x=\omega/T$ with $\omega$ being the photon energy and $T$ the CMB temperature. In this case we can easily find
\begin{equation}\label{final-eq}
B_\textrm{0}=6.76\times 10^{-11}\,\frac{\sqrt{\mu}}{\bar m_a\,C_{a\gamma}}\, \textrm{G},
\end{equation}
where $\bar T$ and $\bar m_a=m_a/eV$ are respectively the resonance temperature and axion mass. Since we are looking for spectral distortion in the redshift interval $2.88\times 10^5\lesssim 1+z\lesssim 2\times 10^6$, the corresponding resonant axion mass is within the interval $2.66\times 10^{-6}\, \textrm{eV}\,\lesssim \bar m_a\lesssim 4.88\times 10^{-5}\, \textrm{eV}$ \cite{Ejlli:2013uda}.

Equation \eqref{heat-B-limit} gives only the contribution to $\mu$ distortion from magnetic field itself. Now we must add to it also the contribution from axion creation from the CMB. Indeed, adding to Eq. \eqref{heat-B-limit}, Eq. \eqref{final-eq} we get the following relation between magnetic field strength, $\mu$-parameter and $\lambda_B$
\begin{equation}\label{small-lambs}
B_0=\sqrt{\mu}\left(1.6\times 10^{-6}\,C_n^{-1/2}(\lambda_B/\lambda_D)^{-(\frac{n+3}{2})}+3.38\times 10^{-11}\frac{1}{\bar m_a\,C_{a\gamma}}\right)\, \textrm{G},\quad (\lambda_B\ll \lambda_D)
\end{equation}
On the other hand, in the limiting case $\lambda_D\ll \lambda_B$ and adding to Eq. \eqref{limit-2}, Eq. \eqref{final-eq} we get
\begin{equation}\label{long-lambs}
B_0=\sqrt{\mu}\left(1.6\times 10^{-6}\,D_n^{-1/2}(\lambda_B/\lambda_D)+3.38\times 10^{-11}\frac{1}{\bar m_a\,C_{a\gamma}}\right)\,\textrm{G},\quad (\lambda_D\ll \lambda_B).
\end{equation}

We notice from Eq. \eqref{long-lambs} that the magnetic field strength depend on the spectral index $n$ only through $D_n$. It is interesting to know at what scales the axion contribution to $\mu$ distortion is smaller than magnetic field contribution. In the case $\lambda_B\ll \lambda_D$ we get
\begin{equation}\label{upper-limit}
\lambda_B\geq \left(4.73\times 10^4\, \frac{\bar m_a C_{a\gamma}}{C_n^{1/2}}\right)^{2/(n+3)}\,\lambda_D,\quad (\lambda_B\ll \lambda_D)
\end{equation}
while in the case $\lambda_D\ll \lambda_B$ we get
\begin{equation}\label{lower-limit}
\lambda_B\leq 2.11\times 10^{-5}\,\frac{D_n^{1/2}}{\bar m_a C_{a\gamma}}\,\lambda_D,\quad (\lambda_D\ll \lambda_B).
\end{equation}
We can see from Eq. \eqref{upper-limit} and Eq. \eqref{lower-limit} that $\lambda_B$ does not depend on the average strength of the magnetic field $B_0$ but only on $n, C_{a\gamma}$ and $\bar m_a$. For example for $n=-2.9, -2, -1, 0, 1, 2, 3$ we have respectively $C_n=1.27, 0.78, 0.77, 1.1, 2.08, 4.93, 14.05$ and $D_n=0.38, 2.71, 4.07, 4.88, 5.43, 5.82, 6.11$. If we consider for example the DFSZ axion model, $n=2$ and axions with mass $\bar m_a=3.5\times 10^{-6}$ eV we have that for $\lambda_B\ll 415.5$ pc, the axion contribution to $\mu$ distortion is subdominant to magnetic field damping for $\lambda_B\geq 172. 5$ pc. In the opposite limit, $\lambda_B\gg 415.5$ pc, we get $\lambda_B\leq 4.03\times 10^{3}$ pc. On the other hand, if we have $\mu$ given by the experiment and ask at what scales axion contribution in the $B_0-\lambda_B$ plane is subdominant to magnetic field damping, we must simply reverse the inequality signs in both Eq. \eqref{upper-limit} and Eq. \eqref{lower-limit}, see Fig. \ref{fig:Fig7}.

In Fig. \ref{fig:Fig1a} the exclusion plot for the scale averaged magnetic field, $B_0$ vs. $\lambda_B$ is shown. In both (a) and (b) the plots for the upper limit on $\mu$ found by COBE \cite{Fixsen:1996nj} are shown. Here we have chosen magnetic fields with $n\geq 2$ which are generated in the early universe by causal mechanisms \cite{Durrer:2003ja}. For such magnetic fields, the field wavelength $\lambda$ or $\lambda_B$ must be smaller than Hubble distance at redshift $z$, namely $\lambda_B\leq H^{-1}(z)$. Indeed, in Fig. \ref{fig:Fig1a} we have chosen $H^{-1}(z_{QCD})\leq\lambda_B\leq H^{-1}(z_\mu)$ where $H^{-1}(z_{QCD})\sim 1$ pc is the QCD comoving horizon and $z_\mu=2.88\times 10^5$ (lower redshift of $\mu$ epoch). Our exclusion and sensitivity plots in Fig. \ref{fig:Fig1a}, Fig. \ref{fig:Fig2a} and Fig. \ref{fig:Fig3a} have been obtained for $\mu$ distortion and $n=2, 3$. The region above the solid line is excluded with no photon-axion mixing taken into account while regions above the dashed, dot dashed and dotted lines are excluded by taking into account it. The exclusion plot with no photon-axion mixing has been obtained by using Eq. \eqref{heat-B-limit} and Eq. \eqref{limit-2} and extrapolating them until $\lambda_B\rightarrow \lambda_D$.

We can see from Fig. \ref{fig:Fig1a}, Fig. \ref{fig:Fig2a} and Fig. \ref{fig:Fig3a} that when we take into account photon-axion mixing, there are significant deviations for 200 pc $\leq\lambda_B\leq$ $10^3$ pc, in comparison with no photon-axion mixing. Depending on the axion mass and axion model, deviations range from a factor 2 until a factor 11. In our plots we have chosen three representative axion masses, $\bar m_a=4.88\times 10^{-5}$ eV, $\bar m_a=1\times 10^{-5}$ eV and $\bar m_a=3.5\times 10^{-6}$ eV. Axions with mass $\bar m_a=4.88\times 10^{-5}$ eV are resonantly produced at the beginning of $\mu$-epoch while axions with mass $\bar m_a\geq 3.5\times 10^{-6}$ eV are experimentally allowed by ADMX collaboration\footnote{To be more precise, ADMX collaboration did not find any axion in the mass range $3.3\mu$ eV-3.5$\mu$ eV.} \cite{ADMX}. Axions with mass $\bar m_a\simeq 3.5\times 10^{-6}$ eV give weaker limits on $B_0$ in comparison with axions with bigger masses. This can be seen from Eq. \eqref{final-eq} where the axion mass is in the denominator.

In the case of expected limits on $\mu$ by future missions such as PIXIE/PRISM \cite{Kogut:2011xw}, our plots are shown as sensitivity plots, see Fig. \ref{fig:Fig2a}, Fig. \ref{fig:Fig3a} and Fig. \ref{fig:Fig5a}. For example we can see in Fig. \ref{fig:Fig2a} that PIXIE/PRISM have a much better sensitivity with respect to COBE in the $B_0-\lambda_B$ plane. Depending on the axion mass and axion model the improvement is in general one or two orders of magnitude. In Fig. \ref{fig:Fig7} plots of Eq. \eqref{final-eq}, Eq. \eqref{small-lambs} and \eqref{long-lambs} are shown. In this figure we can see the regions where the axion contribution in the $B_0-\lambda_B$ plane is dominant or subdominant. In Fig. \ref{fig:Fig8} and Fig. \ref{fig:Fig5a} plots for negative spectral indexes, for example $n=-2.9$ and $n=-2$ are shown. In general magnetic fields with negative spectral indexes are generated by non-casual processes, for example during inflationary epoch \cite{Grasso:2000wj}.

\emph{Summary}. In this work we have considered the impact of spatially varying stochastic magnetic fields and resonant photon-axion mixing on CMB $\mu$ distortion. The contribution of magnetic field to $\mu$ distortion depends on the cut-off scale $\lambda_B$ and on the damping scale $\lambda_D$. On the other hand axion contribution is scale independent as can be seen from Eq. \eqref{final-eq}. Taking into account axion contribution to $\mu$ distortion, in general, one finds weaker limits on the scale averaged magnetic field $B_0$ in comparison with no photon-axion mixing included. Our main results have been shown as exclusion and sensitivity plots in the $B_0-\lambda_B$ plane where the value of the chemical potential has been chosen either equal to the upper limit found by COBE or equal to the expected limit of PIXIE/PRISM.

In this work we have considered only resonant photon-axion mixing on generating a non zero chemical potential. In the resonant case, the axion mass is not arbitrary but is connected with the $\mu$ epoch redshift, $z_\mu$. This constraints the resonant axion mass in the range $2.66\times 10^{-6}\, \textrm{eV}\,\lesssim \bar m_a\lesssim 4.88\times 10^{-5}\, \textrm{eV}$ \cite{Ejlli:2013uda}. Axions with masses outside this interval make non resonant transition into photons.

The inferiority of axion contribution to $\mu$ distortions in comparison with magnetic field damping is scale dependent as given by Eq. \eqref{upper-limit} and Eq. \eqref{lower-limit}. Obviously this is parameter depended and certain number of approximations are in order. If we consider magnetic field generated by casual mechanism, $n\geq 2$ \cite{Durrer:2003ja}, axions with a mass $\bar m_a\geq 3.5\times 10^{-6}$ eV allowed by ADMX \cite{ADMX} and DFSZ axion model we find that bigger contribution to $\mu$ distortion with respect to magnetic field damping occurs for cut-off scales,  $\lambda_B\leq $172.5 pc and $\lambda_B\geq 4\times 10^3$ pc. In the $B_0-\lambda_B$ plane axion contribution dominates over magnetic field damping for scale 172.5 pc $\leq\lambda_B\leq 4\times 10^3$ pc. For scales $\lambda_B\simeq\lambda_D$, $B_0$ weakly depend on the spectral index $n$. For example by using Eq. \eqref{small-lambs} we have that for $\mu< 9\times 10^{-5}$, $\bar m_a=2.66\times 10^{-6}$ eV (lower limit of resonant axion mass) and DFSZ axion, the upper limit on scale averaged magnetic field is $B_0\leq 8.77\times 10^{-8}$ G for $n=2$, $B_0\leq 8.49\times 10^{-8}$ G for $n=3$ and $B_0\leq 9.81\times 10^{-8}$ G for $n=-1$. If we had neglected the contribution of resonant photon-axion contribution to $\mu$ distortion, we would get  $B_0\leq 1.36\times 10^{-8}$ G for $n=2$, $B_0\leq 8\times 10^{-9}$ G for $n=3$ and $B_0\leq 3.45\times 10^{-8}$ G for $n=-1$. Therefore we can conclude that for values of the parameters as assumed above, resonant photon-axion production gives weaker limits on $B_0$ up to a factor in 10 in comparison with no photon-axion mixing included.

Limits on $B_0$ in the case of KSVZ axion model are in general stronger than those from DFSZ axion model (for a given axion mass). For example, if we consider axions with mass $\bar m_a=4.88\times 10^{-5}$ eV (allowed upper limit) and $\mu$ fixed, contribution of resonant photon-axion production to Eq. \eqref{small-lambs} and Eq. \eqref{long-lambs}, is almost marginal. In the $B_0-\lambda_B$ plane the curve corresponding to the KSVZ axion model and $\bar m_a=4.88\times 10^{-5}$ eV is almost indistinguishable from the curve corresponding to the case with no photon-axion mixing included. In general the two curves differs from each other up to a factor 1.5 for 172.5 pc $\leq\lambda_B\leq 4\times 10^3$ pc.

A forecast for the future space missions PIXIE/PRISM has been presented. In Fig. \ref{fig:Fig2a} we can see the level of sensitivity in the case where photon-axion mixing is not included and in the case when it is. In the former case we find that, for scales $\lambda_B\simeq 415.5$ pc, PIXIE/PRISM will probe magnetic fields $B_0\geq 10^{-10}$ G while in the latter case is parameter depended. For axions with masses $\bar m_a\geq 2.66\times 10^{-6}$ eV we find $B_0\geq 2$ nG for DFSZ axion and $B_0\geq 8.8 \times 10^{-10}$ G for KSVZ axion. In the case of axions with masses $\bar m_a=4.88\times 10^{-5}$ eV contribution of photon-axion mixing is marginal as can be seen from  Fig. \ref{fig:Fig2a}. 

It is important to stress two things. First in this work we did not consider any limit from CMB $y$ distortion since at the moment we are currently working on an effective approach of $y$ distortion in the case of photon-axion mixing. Second, contribution of magnetic field damping to $\mu$ distortion has been derived by using linearised magnetohydrodynamic equations in the photon diffusion limit\cite{Jedamzik:1999bm}. Moreover, we have extrapolated both Eq. \eqref{small-lambs} and Eq. \eqref{long-lambs} for $\lambda_B\rightarrow\lambda_D$. A detailed behaviour of $B_0$ around $\lambda_D$ requires numerical integration that is beyond the scope of this paper but we expect that extrapolated results are very accurate for $\lambda_B\rightarrow\lambda_D$.

\vspace{1cm}

 {\bf{AKNOWLEDGMENTS}}:
 This work is supported by POR fellowship of LNGS and by Top 100 program of Novosibirsk State University. The author thanks the hospitality of Novosibirsk State University where this work was initiated.

  \end{document}